\documentstyle[12pt,epsfig]{article}

\catcode`\@=11
\long\def\@makefntext#1{
\protect\noindent \hbox to 3.2pt {\hskip-.9pt  
$^{{\ninerm\@thefnmark}}$\hfil}#1\hfill}		

\def\@makefnmark{\hbox to 0pt{$^{\@thefnmark}$\hss}}  
	
\def\ps@myheadings{\let\@mkboth\@gobbletwo
\def\@oddhead{\hbox{}
\rightmark\hfil\ninerm\thepage}   
\def\@oddfoot{}\def\@evenhead{\ninerm\thepage\hfil
\leftmark\hbox{}}\def\@evenfoot{}
\def\sectionmark##1{}\def\subsectionmark##1{}}

\renewcommand{\thefootnote}{\fnsymbol{footnote}}

\newcounter{sectionc}\newcounter{subsectionc}\newcounter{subsubsectionc}
\renewcommand{\section}[1] {\vspace*{0.6cm}\addtocounter{sectionc}{1} 
\setcounter{subsectionc}{0}\setcounter{subsubsectionc}{0}\noindent 
	{\normalsize\bf\thesectionc. #1}\par\vspace*{0.4cm}}
\renewcommand{\subsection}[1] {\vspace*{0.6cm}\addtocounter{subsectionc}{1} 
	\setcounter{subsubsectionc}{0}\noindent 
	{\normalsize\it\thesectionc.\thesubsectionc. #1}\par\vspace*{0.4cm}}
\renewcommand{\subsubsection}[1]
{\vspace*{0.6cm}\addtocounter{subsubsectionc}{1}
	\noindent {\normalsize\rm\thesectionc.\thesubsectionc.\thesubsubsectionc. 
	#1}\par\vspace*{0.4cm}}

\newcounter{appendixc}
\newcounter{subappendixc}[appendixc]
\newcounter{subsubappendixc}[subappendixc]

\renewcommand{\appendix}[1] {\vspace*{0.6cm}
        \refstepcounter{appendixc}
        \setcounter{figure}{0}
        \setcounter{table}{0}
        \setcounter{equation}{0}
        \renewcommand{\thefigure}{\Alph{appendixc}.\arabic{figure}}
        \renewcommand{\thetable}{\Alph{appendixc}.\arabic{table}}
        \renewcommand{\theappendixc}{\Alph{appendixc}}
        \renewcommand{\theequation}{\Alph{appendixc}.\arabic{equation}}
        \noindent{\bf Appendix \theappendixc #1}\par\vspace*{0.4cm}}

\def\abstracts#1{{
	\centering{\begin{minipage}{12.2truecm}\footnotesize\baselineskip=12pt\noindent
	\centerline{\footnotesize ABSTRACT}\vspace*{0.3cm}
	\parindent=0pt #1
	\end{minipage}}\par}} 

\renewenvironment{thebibliography}[1]
	{\begin{list}{\arabic{enumi}.}
	{\usecounter{enumi}\setlength{\parsep}{0pt}
\setlength{\leftmargin 1.25cm}{\rightmargin 0pt}
	 \setlength{\itemsep}{0pt} \settowidth
	{\labelwidth}{#1.}\sloppy}}{\end{list}}

\newcounter{itemlistc}
\newcounter{romanlistc}
\newcounter{alphlistc}
\newcounter{arabiclistc}

\newcommand{\fcaption}[1]{
        \refstepcounter{figure}
        \setbox\@tempboxa = \hbox{\footnotesize Fig.~\thefigure. #1}
        \ifdim \wd\@tempboxa > 6in
           {\begin{center}
        \parbox{6in}{\footnotesize\baselineskip=12pt Fig.~\thefigure. #1}
            \end{center}}
        \else
             {\begin{center}
             {\footnotesize Fig.~\thefigure. #1}
              \end{center}}
        \fi}

\newcommand{\tcaption}[1]{
        \refstepcounter{table}
        \setbox\@tempboxa = \hbox{\footnotesize Table~\thetable. #1}
        \ifdim \wd\@tempboxa > 6in
           {\begin{center}
        \parbox{6in}{\footnotesize\baselineskip=12pt Table~\thetable. #1}
            \end{center}}
        \else
             {\begin{center}
             {\footnotesize Table~\thetable. #1}
              \end{center}}
        \fi}

\def\@citex[#1]#2{\if@filesw\immediate\write\@auxout
	{\string\citation{#2}}\fi
\def\@citea{}\@cite{\@for\@citeb:=#2\do
	{\@citea\def\@citea{,}\@ifundefined
	{b@\@citeb}{{\bf ?}\@warning
	{Citation `\@citeb' on page \thepage \space undefined}}
	{\csname b@\@citeb\endcsname}}}{#1}}

\newif\if@cghi
\def\cite{\@cghitrue\@ifnextchar [{\@tempswatrue
	\@citex}{\@tempswafalse\@citex[]}}
\def\citelow{\@cghifalse\@ifnextchar [{\@tempswatrue
	\@citex}{\@tempswafalse\@citex[]}}
\def\@cite#1#2{{$\null^{#1}$\if@tempswa\typeout
	{IJCGA warning: optional citation argument 
	ignored: `#2'} \fi}}

 1
 1
 1

\font\ninerm=cmr9

\title{                                                                         
{\vspace{-3cm} \normalsize                                                      
\hfill \parbox{30mm}{DESY 97-034}}\\[25mm]                                      
The Higgs mechanism on the lattice\footnote{\normalsize
Talk given at the Ringberg Workshop: The Higgs puzzle -- what can we
learn from LEP2, LHC, NLC and FMC?, December 1996.} \\[8mm]}
\author{ I. Montvay                                 \\
Deutsches Elektronen-Synchrotron DESY,              \\
Notkestr.\,85, D-22603 Hamburg, Germany}

\date{February, 1997}

\newcommand{\be}{\begin{equation}}   
\newcommand{\ee}{\end{equation}}                                        
\newcommand{\half}{\frac{1}{2}}

\newcommand{\LCB}{\raisebox{-0.3ex}{\mbox{\LARGE$\left\{\right.$}}}
\newcommand{\RCB}{\raisebox{-0.3ex}{\mbox{\LARGE$\left.\right\}$}}}

\begin{document}
\maketitle
\newpage

\centerline{\normalsize\bf THE HIGGS MECHANISM ON THE LATTICE}
\baselineskip=22pt
\centerline{\footnotesize ISTV\'AN MONTVAY}
\baselineskip=13pt
\centerline{\footnotesize\it Deutsches Elektronen-Synchrotron DESY}
\baselineskip=12pt
\centerline{\footnotesize\it Notkestr. 85, D-22603 Hamburg, Germany}
\centerline{\footnotesize E-mail: istvan.montvay@desy.de}

\vspace*{0.9cm}
\abstracts{The lattice regularization of the Higgs sector of the 
 standard model is summarized. The triviality bound and vacuum
 instability bound are described. The question of chiral gauge theories
 is discussed. Some aspects of the numerical simulations of the
 electroweak phase transition are considered.}
 
\normalsize\baselineskip=15pt
\setcounter{footnote}{0}
\renewcommand{\thefootnote}{\alph{footnote}}
\section{Lattice regularization of the Higgs sector}
 The masses of the elementary particles in the standard model are
 due to the Higgs mechanism which is the consequence of the
 spontaneous symmetry breaking in the scalar field sector of the
 theory.
 The spontaneous symmetry breaking is a non-perturbative phenomenon
 which can be well described in the framework of lattice regularization.

\subsection{Lattice action}
 The formulation of the Higgs mechanism on the lattice is well known.
 For this and a general introduction to quantum field theory on the
 lattice see~\cite{MONMUN}.
 The basic ingredients of the Higgs sector in the standard model are
 the Higgs scalar field and the non-abelian SU(2) gauge field of
 weak interactions.
 The Euclidean lattice action of this {\em SU(2) Higgs model} is
 conventionally written as
$$
S[U,\varphi] = \beta \sum_{pl}
\left( 1 - \frac{1}{2} {\rm Tr\,} U_{pl} \right) +
$$
\be \label{eq01}
+ \sum_x \LCB \half{\rm Tr\,}(\varphi_x^+\varphi_x) +
\lambda \left[ \half{\rm Tr\,}(\varphi_x^+\varphi_x) - 1 \right]^2
- \kappa\sum_{\mu=1}^4
{\rm Tr\,}(\varphi^+_{x+\hat{\mu}}U_{x\mu}\varphi_x) \RCB \ .
\ee
 In the first line here we see the Wilson action for the SU(2) gauge
 field, with $U_{x\mu}$ denoting the SU(2) gauge link variable and 
 $U_{pl}$ as the product of four $U$'s around a plaquette.
 In the second line there is the action piece describing the Higgs
 scalar field and its interaction with the gauge field.
 The Higgs field is represented by the complex $2 \otimes 2$ matrix
 in isospin space $\varphi_x$.
 This satisfies the condition $\varphi_x^+ = \tau_2\varphi_x^T\tau_2$
 which ensures that $\varphi_x$ has only four independent real
 components.

 The bare parameters in the lattice action in Eq.~(\ref{eq01}) are:
 $\beta \equiv 4/g^2$ for the gauge coupling $g$, $\lambda$ for the
 scalar quartic coupling and $\kappa$ for the scalar hopping parameter
 related to the bare mass square $\mu_0^2$ by
\be \label{eq02}
\mu_0^2 = (1-2\lambda)\kappa^{-1} - 8 \ .
\ee

 For the understanding of the Higgs mechanism one has to investigate
 the phase structure of the model, first as a function of the bare
 parameters.
 It turns out that there are two ``phases'', namely the
 {\em Higgs-phase} where the Higgs mechanism is operational and the
 {\em confining phase} which is qualitatively similar to QCD with
 scalar quarks.
 There is a phase transition surface
 $\kappa_{cr}=\kappa_{cr}(\beta,\lambda)$ separating the two phases:
 for $\kappa > \kappa_{cr}$ there is the Higgs-phase and for
 $\kappa < \kappa_{cr}$ the confining phase.
 In the standard model the relavant phase is the Higgs-phase, where
 the gauge bosons acquire a non-zero mass due to the non-zero vacuum
 expectation value of the scalar field.

 The phase transition surface is everywhere of first order with a
 discontinuous change of physical quantities, however this surface
 has a boundary at small $\beta$ and large $\lambda$, where the
 phase transition becomes second order with a continuous change
 and infinite correlation lengths.
 Beyond this boundary the two ``phases'' are analytically connected.
 Correspondingly, there exist no (local, gauge invariant) order
 parameters which would distinguish them.
 This is in agreement with Elitzur's theorem on the impossibility
 of spontaneously broken local symmetries~\cite{ELITZUR}.
 In fact, strictly speaking, we are dealing with one single phase on
 the two sides of the transition surface~\cite{FRASHE}.
 The two regions are, however, quantitatively different.

 Since the lattice serves only as a regularization scheme, in order
 to obtain a continuum quantum field theory, one has to remove the
 cut-off, i.~e.~one has to perform the {\em continuum limit}.
 If $a(\kappa,\beta,\lambda)$ denotes the lattice spacing, one
 requires for any physical mass $a \cdot Mass \to 0$.

\subsection{Triviality bound}
 The couplings in the SU(2) Higgs model are not asymptotically free,
 therefore the Gaussian fixed point at zero couplings is not
 appropriate for defining the continuum limit.
 For a non-trivial continuum limit one would need a non-trivial
 ultraviolet fixed point at non-zero couplings.
 This would resolve the problems related to the {\em Landau pole}
 in perturbation theory, which seem to imply that the running quartic
 coupling becomes infinite at high energy scales.

 The non-perturbative lattice investigations show, however, that there
 exists no non-trivial fixed point.
 The consequence is that the renormalized quartic coupling $\lambda_R$
 is zero in the continuum limit:
\be \label{eq03}
\lim_{a \to 0} \lambda_R(\kappa,\beta,\lambda) = 0 \ .
\ee
 The path along which the limit is performed here is quite arbitrary,
 but a usual way to perform continuum limits is along {\em lines of
 constant physics,} with some dimensionless physical quantities fixed.
 The consequence of Eq.~(\ref{eq03}) is that in the SU(2) Higgs sector
 the continuum limit is trivial, that is non-interacting.

 Another, more detailed, formulation of the triviality of the continuum
 limit is that, considering the cut-off theory at non-zero lattice
 spacings ($a \cdot Mass > 0$), there exists a cut-off dependent
 upper bound on the quartic coupling and hence on the ratio
 $M_H/M_W$ ($M_H$ denotes the physical Higgs boson mass, $M_W$ the
 W-boson mass).
 One can show that, as a consequence of renormalizability, the upper
 bound on $\lambda_R$ goes only logarithmically to zero for decreasing
 lattice spacing.
 As a consequence, there exist effective theories with very high
 cut-off's and non-zero renormalized quartic and gauge couplings.
 In these theories with high cut-off's the cut-off effects are
 negligibly small.

 If one tolerates larger cut-off effects the upper bound on the
 Higgs boson mass becomes higher.
 At some point, of course, the cut-off effects become so large that
 the effective theory looses sense.
 In this way one obtains an absolute upper bound~\cite{DASNEU}.
 This absolute upper bound is, of course, somewhat uncertain because
 it depends on ones requirements on the quality of the effective
 theory.
 The results of numerical simulations in the SU(2) Higgs model
 give an absolute upper bound~\cite{LANMON,HASNEU}:
\be \label{eq04}
M_H \leq 9M_W \ .
\ee
 Similar results can also be obtained in the pure scalar $\phi^4$
 model, taking into account the SU(2) gauge coupling only
 perturbatively~\cite{DASNEU,LUSWEI}.
 A recent systematic investigation of the effects of the choice of
 the scalar lattice action gave as a best estimate for the absolute
 upper bound $M_H \leq (710 \pm 60)$ GeV~\cite{HEKLNEVR}.

\subsection{Vacuum instability bound}
\begin{figure}
\begin{center}
\epsfig{file=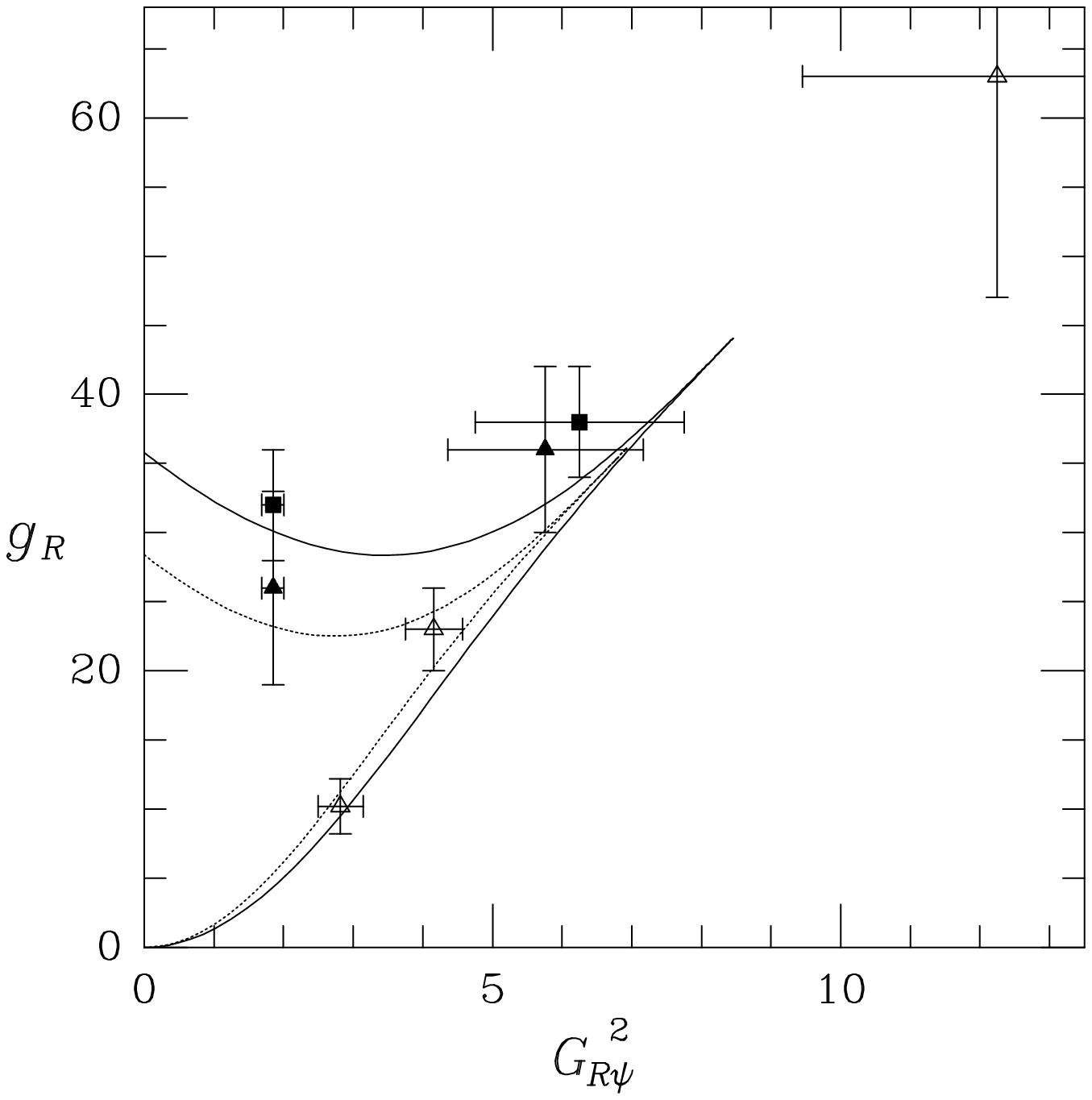,
        width=7.0cm,height=7.0cm,
        angle=0}
\end{center}\vspace{0.5em}
\fcaption{Upper and lower bounds on $g_R$ as a function of
 $G_{R\psi}^2$ together with the one-loop perturbative estimates for 
 cut-off ratios $\Lambda/M_H=3$ (solid curve) and $\Lambda/M_H =4$
 (dotted curve).
 Open points denote the data for the lower bound, whereas full symbols
 are data for the upper bound.
 The $6^3\cdot12$-lattice is represented by triangles, whereas points 
 on $8^3\cdot16$ are denoted by squares.}
\label{fig01}
\end{figure}
 The triviality of couplings remains if Yukawa couplings of the type
 present in the standard model are included. 
 The cut-off dependent upper bounds on the renormalized couplings are 
 such that in phenomenologically relevant models there is no strongly
 interacting Higgs-Yukawa sector, provided that the cut-off scale is
 chosen to be high, say near the scale of grand unification.
 The absolute upper bounds are in general quantitatively similar to
 the upper bounds obtained from the assumption of the unitarity of
 perturbative scattering amplitudes.
 Similarly, the position of the Landau pole in the one-loop 
 $\beta$-function can also be taken as a rough estimate for the upper
 bounds.

 Another cut-off dependent restriction on the renormalized couplings
 originates from the fact that the radiative corrections generated by
 the Yukawa couplings make the renormalized scalar quartic coupling
 larger.
 This obviously implies a lower bound on the Higgs boson mass which
 depends on the strength of the Yukawa couplings.
 In a renormalization group treatment \`a la 
 Coleman-Weinberg~\cite{COLWEI} this manifests itself by the fact that
 the vacuum appears to be unstable at large scales.
 The largest scales in a cut-off quantum field theory are at the
 cut-off, hence the instability occurs at the cut-off.

 It is worth to emphasize that this type of instability is by no means
 a phase transition, which is usually associated to the instability of
 the vacuum in the infrared.
 ``Tunneling'' considerations are in this case out of context because
 they would refer to the possibility of mathematically defining the
 quantum field theory with a cut-off.

 The numerical Monte Carlo simulations in Higgs-Yukawa theories involve
 fermions and are, therefore, much more demanding than the bosonic
 Higgs sector discussed in the previous subsection.
 In addition to the practical difficulties implied by the necessity 
 to evaluate the Grassmannian path integrals, there is also the
 question of principle about the possibility of chiral gauge couplings
 of the fermion fields.
 This will be separately discussed below.

 Neglecting the gauge couplings, there have been several numerical
 studies aiming to determine both the triviality upper bound and
 the vacuum instability lower bound as a function of a strong Yukawa
 coupling.
 For a typical result see Fig.~\ref{fig01} from Ref.~\cite{LIMOMUPLWI}.
 The renormalized quartic coupling is denoted in this figure by $g_R$,
 the renormalized Yukawa coupling by $G_{R\psi}$.
 The numerical data are compared to the perturbative estimates obtained
 from the one-loop $\beta$-function at different cut-off's.

\subsection{No chiral gauge theories?}
 It is an obvious question whether the lattice formulation of the
 Higgs-sector in Eq.~(\ref{eq01}) can also be extended to the full
 standard model?
 From the practical point of view concerning the upper and lower bounds
 discussed above this is not important, because the bounds depend
 essentially only on the strong couplings: the lattice formulation of
 QCD is well known~\cite{MONMUN} and the chiral electroweak
 $\rm SU(2) \otimes U(1)$ couplings are weak.
 Nevertheless, at least as a question of principle, a complete
 non-perturbative formulation of the standard model is obviously
 interesting.
 In fact, this is more than just a question of principle: although in
 most applications the perturbative treatment of the chiral
 electroweak gauge couplings is sufficient, there are some questions
 where the non-abelian non-perturbative nature of the SU(2) coupling
 becomes relevant.
 For instance, this is the case for the electroweak phase transition
 discussed in the next section and also for some instanton induced
 processes in high energy scattering~\cite{RINGESP}.

 It turns out that, in contrast to vector-like gauge theories as QCD
 where the lattice formulation is easy and elegant, the non-perturbative
 lattice formulation of chiral gauge theories seems impossible.
 After many years of struggle one can conjecture that beyond the
 perturbative framework no chiral gauge theories exist!
 The assumptions, which are used as a basis for this conjecture, are:
\begin{itemize}
\item
 quantum field theories are defined as limits of reguralized cut-off
 theories;
\item
 the infinite cut-off limit exists and is independent of the choice of
 the regularization;
\item
 there exists an explicitly gauge invariant local action;
\item
 the formulation can be given by a Euclidean path integral satisfying
 reflection positivity, which implies unitarity after Wick rotation to
 Minkowski space.
\end{itemize}
 Under these, rather natural, assumptions all attempts for constructing
 a chiral gauge theory seem to fail (see, for instance,
 Ref.~\cite{ROMA}).
 Relaxing some of the assumptions allows for some formulations.
 An example is the ``overlap formalism'' in Ref.~\cite{NARNEU}.

 The root of all difficulties is the Nielsen-Ninomiya theorem, which
 tells that under rather general assumptions in a lattice formulation
 there are always {\em fermion doublers} with opposite chirality which
 render the lattice theory non-chiral (vector-like)~\cite{NIENIN}.
 Another characteristic feature of the failure of formulating chiral
 gauge theories on the lattice is the emergence of mirror fermion pairs
 due to the dynamics, even if naively the fermion doublers seem to be
 removed from the physical spectrum.
 There is an intimate relation also to the triviality of the continuum
 limit: the lattice fermion doublers required by the Nielsen-Ninomiya
 theorem cannot be given an arbitrarily large mass because of the
 triviality of the Yukawa couplings.

 The way out of this dilemma is either to give up some of the above
 fundamental assumptions, or to assume the existence of heavy
 {\em mirror leptons and quarks} in nature which render the standard
 model non-chiral at high energy.
 The standard model extended by three mirror pairs of fermion families
 can be formulated without difficulties on the lattice~\cite{MIRROR}.
 The mirror fermions can mix with their normal fermion partners, which 
 imply small breakings of universality of the fermion couplings to 
 gauge bosons~\cite{UNIVER}.
 The mixing also allows for associate production of
 fermion-mirror-fermion pairs at high energy colliders.

\section{Electroweak phase transition}
\begin{figure}
\begin{center}
\epsfig{file=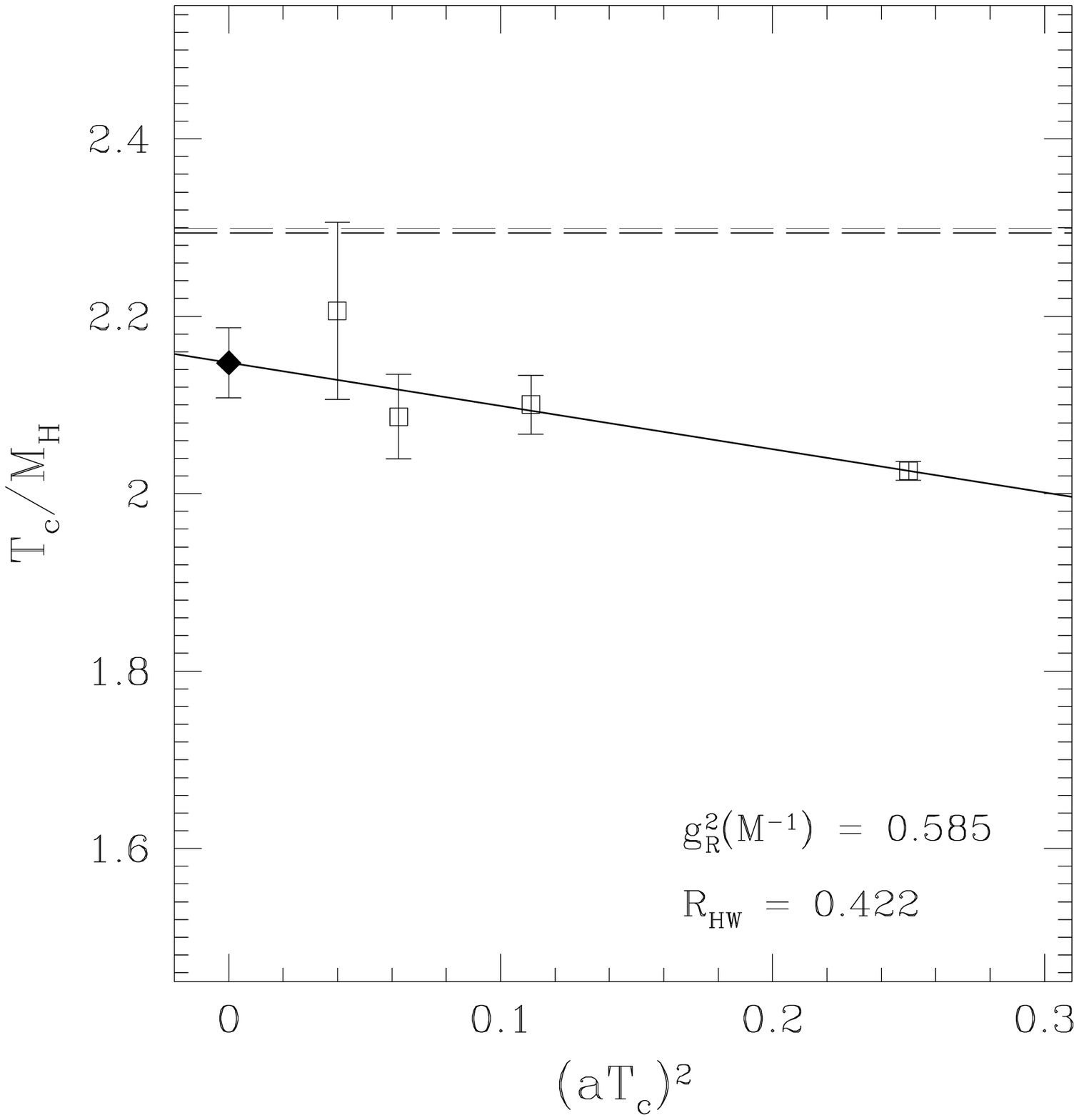,
        width=7.0cm,height=7.0cm,
        bbllx=75pt,bblly=175pt,bburx=600pt,bbury=700pt,angle=0}
\end{center}\vspace{0.5em}
\fcaption{The numerical results for the ratio of the transition
 temperature and the Higgs boson mass $T_c/M_H$ versus
 $(aT_c)^2=L_t^{-2}$.
 The straight line is the extrapolation to very small lattice
 spacings, which gives the continuum value shown by the filled
 symbol. 
 The dashed horizontal lines are the perturbative predictions at order
 $g^3$ (upper) and $g^4$ (lower), respectively.}
\label{fig02}
\end{figure}
 At high temperatures, above the scale of the vacuum expectation value,
 the Higgs mechanism is not operational, the symmetry of the vacuum
 gets restored.
 In the early Universe matter first existed in the symmetry restored 
 confinement phase and went through a phase transition into the
 Higgs-phase.

 The properties of the {\em electroweak phase transition} between the
 confinement and Higgs-phase might have a substantial influence on the
 later history of the Universe.
 The number of baryons is not conserved in the standard model, 
 therefore the small baryon asymmetry of the Universe could perhaps 
 be created in non-equilibrium processes during a strong enough first 
 order electroweak phase transition~\cite{KURUSH}.

\subsection{Four dimensional lattice simulations}
 Near the electroweak phase transition and in the confinement phase
 with restored symmetry infrared singularities render the perturbation
 theory uncertain. 
 Non-perturbative numerical simulations are useful in providing
 numerical control of the perturbative resummation techniques.

 One way of performing the numerical simulations is to first reduce
 the theory by the use of perturbation theory to three dimensions and
 do the numerical simulations there.
 For this approach see the contribution of Shaposhnikov at this
 workshop~\cite{SHAPOSH}.
 It is also possible to perform numerical simulations directly
 in four dimensions, without reduction.
 For some recent results see Refs.~\cite{FOHEJAJAMO,CSFOHEJAMO}.
 In the Higgs boson mass range below 50 GeV the agreement with
 two-loop resummed perturbation theory is reasonably good.
 An example is shown by Fig.~\ref{fig02} from Ref.~\cite{CSFOHEJAMO},
 where the ration of the phase transition temperature to the Higgs
 boson mass $T_c/M_H$ is shown at $M_H \simeq 35$ GeV.
 At higher $M_H$ perturbation theory becomes quite uncertain but
 numerical simulations in four dimensions are still possible and
 informative~\cite{AOKI,CSFOHEHEJAMO}.
\begin{figure}
\begin{center}
\epsfig{file=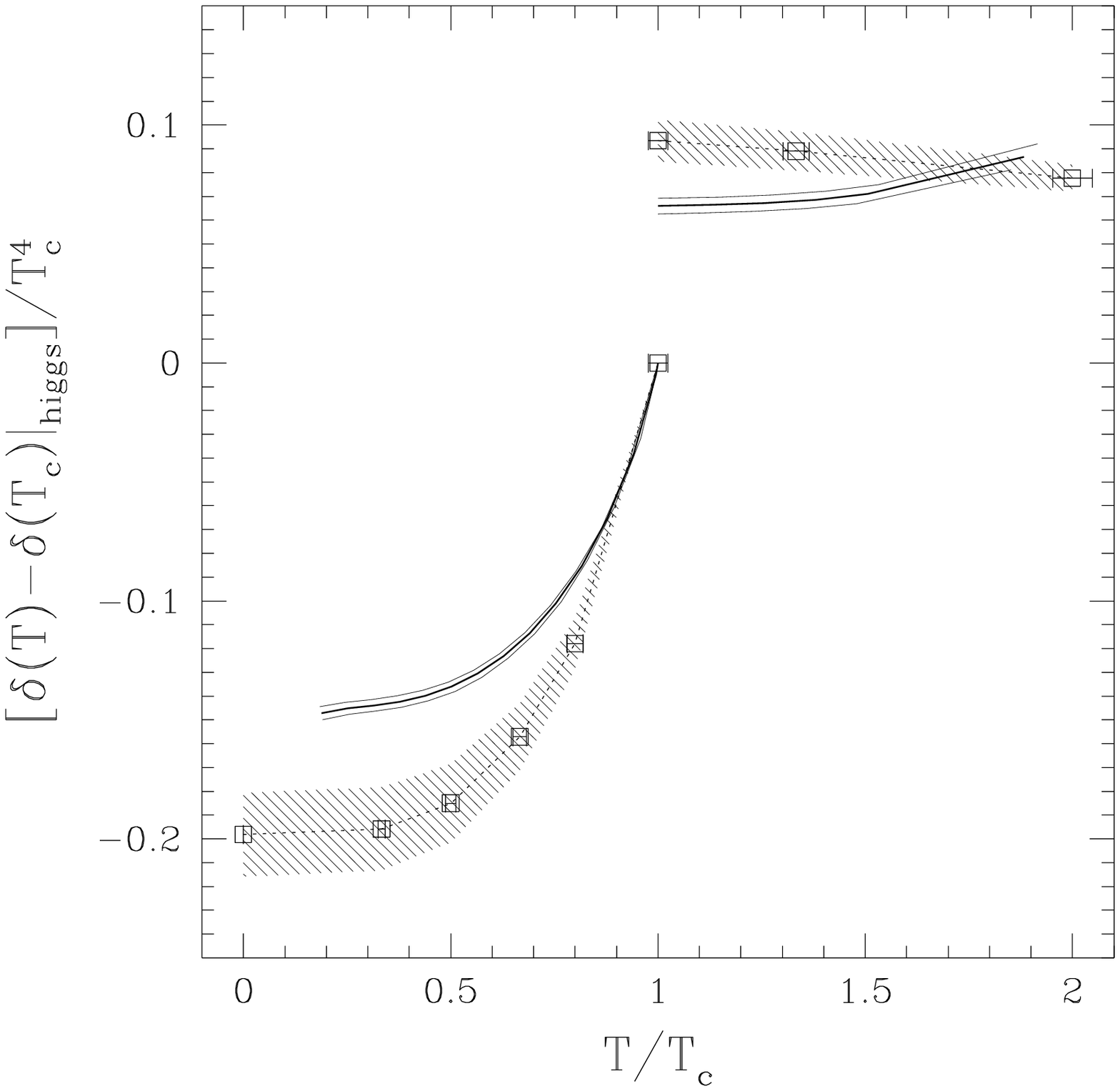,
        width=9.0cm,height=8.0cm,
        bbllx=0pt,bblly=200pt,bburx=600pt,bbury=700pt,angle=0}
\end{center}\vspace{0.0em}
\fcaption{Comparison of the numerical results for 
 $\delta \equiv \epsilon/3 - P$ with those from two-loop
 perturbation theory.
 The shaded areas show the uncertainty of the numerical simulation
 results due to the uncertainty in the derivatives of bare parameters
 along the lines of constant physics.
 The solid lines show the perturbative results together with the
 uncertainties induced by the errors on $R_{HW}$ and $g_R^2$.}
\label{fig03}
\end{figure}

\subsection{A thermodynamical equation of state}
 An important, recently developping, area for numerical simulations is
 the investigation of thermodynamical equations of state in the
 electroweak plasma.
 This is particularly interesting in the high temperature confining
 phase, where perturbation theory is quite uncertain (strictly speaking
 useless), due to the so called ``magnetic mass'' problem.
 A general attitude is to assume a simple ideal gas behaviour.
 Numerical simulations could be very valuable for controlling the
 magnitude of the deviations from this.
 A first example is shown by Fig.~\ref{fig03} from
 Ref.~\cite{CSFOHEJAMO}, where the thermodynamical quantity
 $\delta \equiv \epsilon/3 - P$ is depicted ($P$ denotes the pressure,
 $\epsilon$ the energy density).
 As one can see, the deviation from the ideal gas behaviour $\delta=0$
 is small in the confining phase.
 The relevant ratio $\delta/T^4=\delta/T_c^4 \cdot T_c^4/T^4$ is in
 fact non-zero mainly near the phase transion temperature
 $T = T_c$.



\section{References}
\vspace{-1em}
\begin{thebibliography}{99}
%
\bibitem{MONMUN} 
I. Montvay, G. M\"unster, 
{\it Quantum Fields on a Lattice} 
(Cambridge University Press, Cambridge, 1994)
%
\bibitem{ELITZUR}
S. Elitzur,
{\it Phys. Rev.} {\bf D12} (1975) 3978.
%
\bibitem{FRASHE}
E. Fradkin, S. Shenker,
{\it Phys. Rev.} {\bf D19} (1979) 3682.
%
\bibitem{DASNEU}
R. Dashen, H. Neuberger,
{\it Phys. Rev. Lett.} {\bf 50} (1983) 1897.
%
\bibitem{LANMON}
W. Langguth, I. Montvay,
{\it Z. Phys.} {\bf C36} (1987) 725.
%
\bibitem{HASNEU}
A. Hasenfratz, T. Neuhaus,
{\it Nucl. Phys.} {\bf B297} (1988) 205.
%
\bibitem{LUSWEI}
M. L\"uscher, P. Weisz,
{\it Nucl. Phys.} {\bf B318} (1989) 705.
%
\bibitem{HEKLNEVR}
U.M. Heller, M. Klomfass, H. Neuberger, P. Vranas,
{\it Nucl. Phys.} {\bf B405} (1993) 555.
%
\bibitem{COLWEI}
S. Coleman, E. Weinberg,
{\it Phys. Rev.} {\bf D7} (1973) 1888.
%
\bibitem{LIMOMUPLWI}
L. Lin, I. Montvay, G. M\"unster, M. Plagge, H. Wittig,
{\it Phys. Lett.} {\bf B317} (1993) 143. 
%
\bibitem{RINGESP}
A. Ringwald,
{\it Nucl. Phys.} {\bf B330} (1990) 1; \\
O. Espinosa,
{\it Nucl. Phys.} {\bf B343} (1990) 310. 
%
\bibitem{ROMA}
{\it Nonperturbative Aspects of Chiral Gauge Theories,} Rome, 1992;
published in {\it Nucl. Phys. Proc. Suppl.} {\bf 29BC} (1992).
%
\bibitem{NARNEU}
R. Narayanan, H. Neuberger,
{\it Nucl. Phys.} {\bf B443} (1995) 305.
%
\bibitem{NIENIN}
H.B. Nielsen and H. Ninomiya,
{\it Nucl. Phys.} {\bf B185} (1981) 20; 
erratum ibid. {\bf B195} (1981) 541;
{\it Nucl. Phys.} {\bf B193} (1981) 173.
%
\bibitem{MIRROR}
I. Montvay,
{\it Phys. Lett.} {\bf 199B} (1987) 89;
{\it Nucl. Phys. Proc. Suppl.} {\bf 4} (1988) 443;
{\it Phys. Lett.} {\bf B205} (1988) 315;  
{\it Nucl. Phys. Proc. Suppl.} {\bf 29BC} (1992) 159.
%
\bibitem{UNIVER}
I. Montvay,
{\it Phys. Lett.} {\bf B382} (1996) 104.
%
\bibitem{KURUSH}                                                                
V.A. Kuzmin, V.A. Rubakov, M.E. Shaposhnikov,                                   
{\it Phys. Lett.} {\bf B155} (1985) 36.
%
\bibitem{SHAPOSH}
M.E. Shaposhnikov, these Proceedings.
%
\bibitem{FOHEJAJAMO}
Z. Fodor, J. Hein, K. Jansen, A. Jaster, I. Montvay,
{\it Nucl. Phys.}  {\bf B439} (1995) 147.
%
\bibitem{CSFOHEJAMO}
F. Csikor, Z. Fodor, J. Hein, A. Jaster, I. Montvay,
{\it Nucl. Phys.} {\bf B474} (1996) 421.
%
\bibitem{AOKI} 
Y. Aoki,
{\it Nucl. Phys. Proc. Suppl.} {\bf 53} (1997) 609;
preprint hep-lat/9612023.
%
\bibitem{CSFOHEHEJAMO}
F. Csikor, Z. Fodor, J. Hein, J. Heitger, A. Jaster, I. Montvay,
{\it Nucl. Phys. Proc. Suppl.} {\bf 53} (1997) 612.
%
\end{thebibliography}
\end{document}